\documentclass[twocolumn,fleqn]{svjour2}    % twocolumn
\smartqed  % flush right qed marks, e.g. at end of proof
\usepackage{graphicx}
\begin{document}

\title{An approach to the Riemann problem for SPH inviscid ideal flows}
\subtitle{a reformulation for the state equation}

\titlerunning{EoS and the Riemann problem for ideal flows}        % if too long for running head

\author{G. Lanzafame}


\institute{G. Lanzafame \at
              INAF - Osservatorio Astrofisico di Catania, Via S. Sofia
              78 - 95123 Catania, Italy \\
              Tel.: +39-0957332316\\
              Fax: +39-095330592\\
              \email{glanzafame@oact.inaf.it}           %  \\
}

\date{Received: date / Accepted: date}

\maketitle

\begin{abstract}
  In the physically non viscous fluid dynamics, "shock capturing" methods adopt either an artificial viscosity contribution or an appropriate Riemann solver algorithm. These techniques are necessary to solve the strictly hyperbolic Euler equations if flow discontinuities (the Riemann problem) must be solved. A necessary dissipation is normally used in such cases. An explicit artificial viscosity contribution is normally adopted to smooth out spurious heating and to treat transport phenomena. Such a treatment of inviscid flows is also widely adopted in the Smooth Particle Hydrodynamics (SPH) finite volume free Lagrangian scheme. In other cases, the intrinsic dissipation of Godunov - type methods is implicitly useful. Instead "shock tracking" methods normally use the Rankine - Hugoniot jump conditions to solve such problem. A simple, effective solution of the Riemann problem in inviscid ideal gases is here proposed, based on an empirical reformulation of the equation of state (EoS) in the Euler equations in fluid dynamics, whose limit for a motionless gas coincides with the classical EoS of ideal gases. The application of such effective solution of the Riemann problem excludes any dependence, in the transport phenomena, on particle resolution length $h$ in non viscous SPH flows. Results on 1D shock tube tests are here shown.

\end{abstract}

\section{Introduction}
\label{intro}

  In both Lagrangian and Eulerian inviscid fluid dynamics, a small dissipation is normally useful to handle discontinuities in the flow (the Riemann problem). An artificial viscosity is introduced in SPH, as a shock capturing method, to prevent particle interpenetration and to smooth out spurious heating in the flow to solve the strictly hyperbolic system of Euler equations. The introduction of such a small dissipation, to solve the Euler equations, is also currently adopted to produce both mass and angular momentum transport in SPH physically inviscid modelling of accretion discs in astrophysics (Molteni et al. 1991, 1994; Lanzafame et al. 1992, 1993, 1994, 2000, 2001; Belvedere et al. 1993; Chakrabarti \& Molteni 1993; Meglicki et al. 1993; Murray 1996; Lanzafame \& Belvedere 1997, 1998; Okazaki et al. 2002). Efforts were accomplished in SPH to solve both the "approximate" and the "exact" Riemann problem, either via a reformulation of the artificial viscosity term (Balsara 1995; Monaghan 1997; Morris \& Monaghan 1997) or via Godunov techniques (Yukawa et al. 1997; Inutsuka 2002; Parshikov \& Medin 2002 Molteni \& Bilello 2003) which also include an "intrinsic" dissipation. In the first case, a reformulation of the artificial viscosity could be necessary because, for "weak shocks" or low Mach numbers, the fluid becomes "too viscous" and angular momentum and vorticity could be non - physically transferred. A technique to solve the Riemann problem, based on a reformulation of the EoS in the Euler equations, is here presented, where particle SPH pressure terms are recalculated without any artificial viscosity contribution. Since shock flows are non-equilibrium events of fluid dynamics, we pay attention to the fact that the EoS: $p = (\gamma - 1) \rho \epsilon$ for ideal flows cannot exactly be applied to solve the Riemann problem. In fact such equation is strictly valid only for equilibrium or for quasi-equilibrium thermodynamic state, when the velocity of propagation of perturbations equals the sound velocity. Successful results, based on some form of mathematical dissipation introduced within the strictly hyperbolic system of Euler equations for ideal flows, are obtained due to the necessity to correct such congenital defect. In \S 2 of this paper, we briefly remind how does SPH method work for ideal non viscous flows. In the same \S 2 we also outline which evolution has been accomplished in the explicit artificial viscosity dissipation description. Instead, in \S 3, we show how to reformulate the EoS, according to the Riemann problem for inviscid, ideal gases. Applications to 1D shock tubes (Sod 1978), to solve shocks are also shown in \S 4.

%
%__________________________________________________________________
%
%

\section{The Euler equations and their SPH formulation}
\label{sec:1}

\subsection{SPH and artificial viscosity for non - viscous flows}

  As for Lagrangian ideal non - viscous gas hydrodynamics, the relevant equations (Euler equations) are:

\begin{equation}
\frac{d\rho}{dt} + \rho \nabla \cdot \underline{v} = 0 \hfill \mbox{continuity equation}
\end{equation}

\begin{equation}
\frac{d \underline{v}}{dt} = - \frac{\nabla p}{\rho} \ \ \ \ \ \ \ \hfill \mbox{momentum equation}
\end{equation}

\begin{equation}
\frac{d \epsilon}{dt} = - \frac{p}{\rho} \nabla \cdot \underline{v} \hfill \mbox{energy equation}
\end{equation}

\begin{equation}
p = (\gamma - 1) \rho \epsilon \hfill \mbox{perfect gas equation}
\end{equation}

\begin{equation}
\frac{d \underline{r}}{dt} = \underline{v} \hfill \mbox{kinematic equation}
\end{equation}

  The most of the adopted symbols have the usual meaning: $d/dt$ stands for the Lagrangian derivative, $\rho$ is the gas density, $\epsilon$ is the thermal energy per unit mass. The adiabatic index $\gamma$ has the meaning of a numerical parameter whose value lies in the range between $1$ and $5/3$, in principle.

 The SPH method is a Lagrangian scheme that discretizes the fluid into moving interacting and interpolating domains called "particles". All particles move according to pressure and body forces. The method makes use of a Kernel $W$ useful to interpolate a physical quantity $A(\underline{r})$ related to a gas particle at position $\underline{r}$ according to (Monaghan 1986, 1992):

\begin{equation}
A(\underline{r}) = \int_{D} A(\underline{r}') W(\underline{r}, \underline{r}', h) d \underline{r}'
\end{equation}

$W(\underline{r}, \underline{r}', h)$, the interpolation Kernel, is a continuous function - or two connecting continuous functions whose derivatives are continuous even at the connecting point - defined in the spatial range $2h$, whose limit for $h \rightarrow 0$ is the Dirac delta distribution function. All physical quantities are described as extensive properties smoothly distributed in space and computed by interpolation at $\underline{r}$. In SPH terms we write:

\begin{equation}
A_{i} = \sum_{j=1}^{N} \frac{A_{j}}{n_{j}} W(\underline{r}_{i}, \underline{r}_{j}, h) = \sum_{j=1}^{N} \frac{A_{j}}{n_{j}} W_{ij}
\end{equation}

where the sum is extended to all particles included within the interpolation domain $D$, $n_{j} = \rho_{j}/m_{j}$ is the number density relative to the jth particle. $W(\underline{r}_{i}, \underline{r}_{j}, h) \leq 1$ is the adopted interpolation Kernel whose value is determined by the relative distance between particles $i$ and $j$. Typically, such cubic spline Kernels $W(r_{ij},h)$ are in the form:

\begin{equation}
W_{ij} = \frac{1}{\pi h^{3}} \left\{ \begin{array}{ll}
1 - \frac{3}{2} r^{2}_{ij} + \frac{3}{4} r^{3}_{ij} & \textrm{if $0 \leq r_{ij} \leq 1$}\\
\frac{1}{4} (2 - r_{ij})^{3} & \textrm{if $1 \leq r_{ij} \leq 2$}\\
0 & \textrm{otherwise,}
\end{array} \right.
\end{equation}

even though also Gaussian form are adopted. In expression (8), $r_{ij} = |\underline{r}_{i} - \underline{r}_{j}|/h$ represents the module of the radial distance between particles $i$ and $j$ in units of $h$.

  In SPH formalism, equations (2) and (3) take the form:
  
\begin{equation}
\frac{d \underline{v}_{i}}{dt} = - \sum_{j=1}^{N} m_{j} 
\left( \frac{p_{i}}{\rho_{i}^{2}} + \frac{p_{j}}{\rho_{j}^{2}} \right) \nabla_{i} W_{ij}
\end{equation}

\begin{equation}
\frac{d \epsilon_{i}}{dt} = \frac{1}{2} \sum_{j=1}^{N} m_{j} \left( \frac{p_{i} }{\rho_{i}^{2}} + \frac{p_{j}}{\rho_{j}^{2}}\right)  \underline{v}_{ij} \cdot \nabla_{i} W_{ij}
\end{equation}

where $\underline{v}_{ij} = \underline{v}_{i} - \underline{v}_{j}$ and $m_{j}$ is the mass of jth particle.

  For a better energy conservation, the total energy $E = (\epsilon + \frac{1}{2} v^{2})$ can also be introduced in the SPH formulation:

\begin{equation}
\frac{d}{dt} E_{i} = - \sum_{j=1}^{N} m_{j} \left( \frac{p_{i} \underline{v}_{i}}{\rho_{i}^{2}} + \frac{p_{j} \underline{v}_{j}}{\rho_{j}^{2}} \right) \cdot \nabla_{i} W_{ij}
\end{equation}

of the energy equation:

\begin{equation}
\frac{d}{dt} \left( \epsilon + \frac{1}{2} v^{2}\right) = - \frac{1}{\rho} \nabla \cdot \left( p \underline{v} \right).
\end{equation}

In this scheme the continuity equation takes the form:

\begin{equation}
\frac{d\rho_{i}}{dt} = \sum_{j=1}^{N} m_{j} \underline{v}_{ij} \cdot 
\nabla_{i} W_{ij}
\end{equation}

or, as we adopt, it can be written as:

\begin{equation}
\rho_{i} = \sum_{j=1}^{N} m_{j} W_{ij}
\end{equation}

which identifies the natural space interpolation of particle densities according to equation (7).

  The pressure term includes the artificial viscosity contribution given by Monaghan (1985, 1992) and Monaghan \& Lattanzio (1985), with an appropriate thermal diffusion term which reduces shock fluctuations. It is given by:

\begin{equation}
\eta_{ij} = \alpha \mu_{ij} + \beta \mu_{ij}^{2},
\end{equation}

where

\begin{equation}
\mu_{ij} = \left\{ \begin{array}{ll}
\frac{2 h \underline{v}_{ij} \cdot \underline{r}_{ij}}{(c_{si} + c_{sj}) (r_{ij}^{2} + \xi^{2})} & \textrm{if $\underline{v}_{ij} \cdot \underline{r}_{ij} < 0$}\\
\\
0 & \textrm{otherwise}
\end{array} \right.
\end{equation}

with $c_{si}$ being the sound speed of the ith particle, $\underline{r}_{ij} = \underline{r}_{i} - \underline{r}_{j}$, $\xi^{2} \ll h^{2}$, $\alpha \approx 1$ and $\beta \approx 2$. These $\alpha$ and $\beta$ parameters of the order of the unity (Lattanzio et al. 1985) are usually adopted to damp oscillations past high Mach number shock fronts developed by non-linear instabilities (Boris \& Book 1973). SPH method, like other finite difference schemes, is far from the continuum limit. The linear $\alpha$ term is based on the viscosity of a gas. The quadratic ($\beta$, Von Neumann-Richtmyer-like) artificial viscosity term is necessary to handle strong shocks. Linear $\alpha$ and quadratic $\beta$ artificial viscosity terms (usually $\sim 1$ and sometimes, in some specific cases, $< 1$) are chosen $= 1$ and $= 2$, respectively. In the physically inviscid SPH gas dynamics, angular momentum transport is mainly due to the artificial viscosity included in the pressure terms as:

\begin{equation}
\frac{p_{i}}{\rho_{i}^{2}} + \frac{p_{j}}{\rho_{j}^{2}} = \left( \frac{p_{i}}{\rho_{i}^{2}} + \frac{p_{j}}{\rho_{j}^{2}} \right)_{gas} (1 + \eta_{ij})
\end{equation}

where terms into parentheses refer to intrinsic gas properties.

  In SPH conversion (eqs. 9, 10, 11, 13, 14) of mathematical equations (eqs. 1 to eq. 4) there are two principles embedded. Each SPH particle is an extended, spherically symmetric domain where any physical quantity $f$ has a density profile $f W(\underline{r}_{i}, \underline{r}_{j}, h) \equiv f W(|\underline{r}_{i} - \underline{r}_{j}|, h)$. Besides, the fluid quantity $f$ at the position of each SPH particle could be interpreted by filtering the particle data for $f(\underline{r})$ with a single windowing function whose width is $h$. So doing, fluid data are considered isotropically smoothed all around each particle along a length scale $h$. Therefore, according to such two concepts, the SPH value of the physical quantity $f$ is both the superposition of the overlapping extended profiles of all particles and the overlapping of the closest smooth density profiles of $f$.

\subsection{Progresses on artificial viscosity formulation}

  Strong shock require $\alpha = 1$. However for weak shocks and for small Mach number flows, the fluid becomes "too viscous" and both angular momentum and vorticity are transferred unphysically. To solve such problem, several solutions are proposed:

  a) the formulation of a "limiter" (Balsara 1995), multiplying the artificial viscosity terms $\eta_{i}$ for $f_{ij} = 0.5 (f_{i} + f_{j})$, where

\begin{equation}
f_{i} = \frac{|\nabla \cdot \underline{v_{i}}|}{|\nabla \cdot \underline{v_{i}}| + |\nabla \times \underline{v_{i}}| + 10^{-4} c_{si}/h}.
\end{equation}

  It reduces the unhysical spread of angular momentum in whirling flows up to $20$ times. It is $\approx 1$ for planar shocks, while it increases if the tangential kinematics is relevant;

  b) a switch to reduce artificial viscosity (Morris \& Monaghan 1997). In this hypothesis, for each $ith$ particle the $\alpha$ evolves according to a decay equation:

\begin{equation}
\frac{d \alpha_{i}}{dt} = - \frac{\alpha_{i} - \alpha^{\ast}}{\tau_{i}} + S_{i},
\end{equation}

where $\alpha^{\ast} \approx 0.1$, $\tau_{i} = h/c_{s}$ and the source term $S_{i} = f_{i} \max(- \nabla \cdot \underline{v}_{i}, 0)$.

  c) a reformulation of the artificial viscosity according to the Riemann problem (Monaghan 1997). Particles $i$ and $j$ are considered as the equivalent left "$l$" and right "$r$" states of a given contact interface. The 1D Riemann problem is taken into account along the line joining them. Being the Euler equations in conservative form:

\begin{equation}
\frac{\partial \underline{s}}{\partial t} + \frac{\partial \underline{f}}{\partial x} = 0,
\end{equation}

the simplest Euler technique of integration is:

\begin{equation}
\underline{s}_{i}^{n+1} = \underline{s}_{i}^{n} - \frac{\Delta t}{\Delta x} \left[ \underline{f}^{\ast}(\underline{s}_{i}, \underline{s}_{i-1}) - \underline{f}^{\ast}(\underline{s}_{i-1}, \underline{s}_{i}) \right],
\end{equation}

where numerical fluxes (Mart\'i et al. 1991)

\begin{equation}
\underline{f}^{\ast}(\underline{s}_{l}, \underline{s}_{r}) = 0.5 \left( \underline{f}_{l}^{\ast} + \underline{f}_{r}^{\ast} - \sum_{j=1}^{3} |\lambda_{j}^{\ast} \Delta \omega_{j}^{\ast} \underline{e}_{j}| \right),
\end{equation}

where the $\underline{e}_{j}$ are the eigenvectors of the Jacobian matrix $\boldmath{A} = \partial \underline{f}/\partial \underline{s}$ and $\lambda_{j}^{\ast}$ is an average of $\lambda$ for the "$l$" and "$r$" states. $\Delta \omega_{j}^{\ast}$ are the "jumps" of $\underline{s}$ across the characteristics:

\begin{equation}
\underline{s}_{r} - \underline{s}_{l} = \sum_{j=1}^{3} \Delta \omega_{j}^{\ast} \underline{e}_{j}.
\end{equation}

  For 1D ideal flows, the eigenvalues are $v$, $v + c_{s}$ and $v - c_{s}$, where the two including the sound velocity are the velocities of propagation of sound referred to the frame whose fluid velocity is $v$. Assuming that the jump in the velocity across characteristics could physically be $\underline{v}_{ij} \cdot \underline{r}_{ij}/|\underline{r}_{ij}|$ and that a signal velocity $v_{sig}$ corresponds to the above eigenvalues, $|\lambda_{j}^{\ast}| \Delta \omega_{j}^{\ast}$ corresponds to $v_{sig, ij} \underline{v}_{ij} \cdot \underline{r}_{ij}/|\underline{r}_{ij}|$. So doing, the artificial pressure contribution in the momentum equation is:

\begin{equation}
\left( \frac{p_{i}}{\rho_{i}^{2}} + \frac{p_{j}}{\rho_{j}^{2}} \right)_{gas}  \eta_{ij} = - \frac{K v_{sig, ij} \underline{v}_{ij} \cdot \underline{r}_{ij}/|\underline{r}_{ij}|}{\rho_{ij}},
\end{equation}

where $K$ is an arbitrary constant $\approx 1$ and $\rho_{ij} = 0.5 (\rho_{i} + \rho_{j})$.

  As far as the energy equation is concerned, the SPH formulation in this case is:
  
\begin{eqnarray}
\frac{d}{dt} E_{i} & = & - \sum_{j=1}^{N} m_{j} \left( \frac{p_{i} \underline{v}_{i}}{\rho_{i}^{2}} + \frac{p_{j} \underline{v}_{j}}{\rho_{j}^{2}} \right)_{gas} \cdot \nabla_{i} W_{ij}  \nonumber \\
& & - \frac{K v_{sig, ij} e_{ij}^{\ast} \underline{r}_{ij}/|\underline{r}_{ij}|}{\rho_{ij}} \cdot \nabla_{i} W_{ij},
\end{eqnarray}

where $e_{ij}^{\ast} = e_{i}^{\ast} - e_{j}^{\ast}$ and $e_{i}^{\ast} = 0.5 (\underline{v}_{i} \cdot \underline{r}_{ij}/|\underline{r}_{ij}|)^{2} + \epsilon_{i}$.

  The signal velocities for the 1D Lagrangian Riemann problem on ideal flows are reported in (Whitehurst 1995), on the base of Gottlieb \& Groth (1988) and Toro (1992) results. The pressure on the contact interface $p^{\ast}$ links the left and the right states. For systems with one shock and one rarefaction wave $p^{\ast} \in [p_{l}, p_{r}]$ and

\begin{equation}
p^{\ast} = \left[ \frac{c_{sl} + c_{sr} + (v_{l} - v_{r}) (\gamma - 1)/2}{c_{sl}/p_{l}^{(\gamma - 1)/2 \gamma} + c_{sr}/p_{r}^{(\gamma - 1)/2 \gamma}} \right]^{2 \gamma/(\gamma - 1)}.
\end{equation}

In the case of two shocks, a more complicated relation (Gottlieb \& Groth 1988) also involves the velocity $v^{\ast}$ on the contact interface. However, for some practical purposes, $p^{\ast} = \max(p_{l}, p_{r})$ and $v^{\ast} = 0.5 (v_{l} + v_{r})$. Notice that in the presence of two very strong rarafaction waves $p^{\ast} = 0$.

  The Lagrangian two velocities of waves spawned by the interface are:

\begin{equation}
\lambda_{l} = \left\{ \begin{array}{ll}
v_{l} - c_{sl} [1 + \frac{(\gamma - 1) (p^{\ast}/p_{l} - 1)}{2 \gamma}] & \textrm{if $p^{\ast}/p_{l} > 1$}\\
\\
v_{l} - c_{sl} & \textrm{if $p^{\ast}/p_{l} \leq 1$}
\end{array} \right.
\end{equation}

and

\begin{equation}
\lambda_{r} = \left\{ \begin{array}{ll}
v_{r} + c_{sr} [1 + \frac{(\gamma - 1) (p^{\ast}/p_{r} - 1)}{2 \gamma}] & \textrm{if $p^{\ast}/p_{r} > 1$}\\
\\
v_{r} + c_{sr} & \textrm{if $p^{\ast}/p_{r} \leq 1$}
\end{array} \right.
\end{equation}

In the Lagrangian description, being $v_{sig}$ the speed of propagation of perturbation from $i$ to $j$ and vice-versa ($l \leftrightarrow r$),

\begin{eqnarray}
v_{sig, i \rightarrow j} & = & c_{si} - \underline{v}_{i} \cdot \underline{r}_{ij}/|\underline{r}_{ij}| \\
v_{sig, j \rightarrow i} & = & - c_{sj} - \underline{v}_{j} \cdot \underline{r}_{ij}/|\underline{r}_{ij}| \\
v_{sig} & = & v_{sig, i \rightarrow j} - v_{sig, j \rightarrow i} \\
& = & c_{si} + c_{sj} - \underline{v}_{ij} \cdot \underline{r}_{ij}/|\underline{r}_{ij}|
\end{eqnarray}

having considered the versus going from $i$ to $j$. By performing some numerical checks, Monaghan (1997) also suggested other similar algebraic expressions.

\subsection{Some cautions in using artificial viscosities}

  In Shu (1992), some cautionary remarks are reported on the adoption of artificial dissipation. In particular, "Because the magnitude of the viscous term does not affect the net shock jump conditions, many numerical schemes implicitly or explicitly incorporate the trick of {\it artificial viscosity} for halting the ever-growing steepening tendency produced by nonlinear effects, thereby gaining the automatic insertion of shock waves wherever they are needed (in time-dependent calculations)." However, the numerical viscous term should be large enough to spread out shock transitions only over a few resolution length, making shock waves resolvable. In this sense, any mathematical dissipation should be considered a useful mathematical "trick" to get correct results.

\section{How EoS matches the Riemann problem}

  The solution of the Riemann problem is obtained at the interparticle contact points among particles, where a pressure and a velocity, relative to the flow discontinuity, are computed. This is also clearly shown in Inutsuka (2002); Parshikov \& Medin (2002); Molteni \& Bilello (2993), where the new pressure $p^{\ast}$ and velocity $v^{\ast}$ are reintroduced in the Euler equations instead of $p$ and $v$ to obtain the new solutions compatible with inviscid flow discontinuities. In SPH, we pay attention in particular (Inutsuka 2002; Parshikov \& Medin 2002) to the particle pressures $p_{i}$ and $p_{j}$, in the SPH formulation of the momentum and energy equations (8) and (9), whose substitution with pressures $p_{i}^{\ast}$ and $p_{j}^{\ast}$, solutions of the Riemann problem, excludes any artificial viscosity adoption, although a dissipation, implicitly introduced, is necessary. Therefore, the solution of the Lagrangian Riemann problem, only in its pressure computation, is enough to interface SPH with the Riemann problem solution.

  The key concept is that the physical interpretation of the application of the artificial viscosity contribution in the pressure terms corresponds to a reformulation of the EoS for inviscid ideal gases, whose equation:

\begin{equation}
p|_{gas} = (\gamma - 1) \rho \epsilon
\end{equation}

is strictly applied in fluid dynamics when the gas components do not collide with each other. In the case of gas collisions, it modifies in:

\begin{equation}
p^{\star} = (\gamma - 1) \rho \epsilon + \textrm{other.}
\end{equation}

  The further term takes into account the velocity of perturbation propagation (Monaghan 1997). This velocity equals the ideal gas sound velocity $c_{s}$ whenever we treat non collisional gases in equilibrium or in the case of rarefaction waves. Instead, it includes the "compression velocity": $\underline{v}_{ij} \cdot \underline{r}_{ij}/|\underline{r}_{ij}|$ in the case of shocks (eq. 29). In the first case, we write the EoS for inviscid ideal gases as:

\begin{equation}
p|_{gas} = \frac{\rho}{\gamma} c_{s}^{2},
\end{equation}

where $c_{s} = (\gamma p/\rho)^{1/2} = [\gamma (\gamma - 1) \epsilon]^{1/2}$. Instead, in the second case, the new formulation for the EoS is obtained squaring $v_{sig, i \rightarrow j}$, so that $c_{s}^{2} (1 - v_{shock}/c_{s})^{2}$ is an energy per unit mass in the case of shock. Hence:

\begin{equation}
p^{\star} = \left\{ \begin{array}{ll}
\frac{\rho}{\gamma} c_{s}^{2} \left(1 - \frac{v_{shock}}{c_{s}} \right)^{2} & \textrm{if $v_{shock} < 0$}\\
\\
\frac{\rho}{\gamma} c_{s}^{2} & \textrm{if $v_{shock} \ge 0$}
\end{array} \right.
\end{equation}

  In the SPH scheme, being:

\begin{equation}
p_{i}^{\star} = \frac{\rho_{i}}{\gamma} c_{si}^{2} \left(1 - \frac{v_{shock,i}}{c_{si}} \right)^{2},
\end{equation}

\begin{equation}
v_{shock,i} = \left\{ \begin{array}{ll}
\frac{\underline{v}_{ij} \cdot \underline{r}_{ij}}{|\underline{r}_{ij}|} & \textrm{if $\underline{v}_{ij} \cdot \underline{r}_{ij} < 0$}\\
\\
0 & \textrm{otherwise.}
\end{array} \right.
\end{equation}

  This formulation introduces the "shock pressure term" $\rho (v_{shock}^{2} - 2 v_{shock} c_{s})/\gamma$, whose linear and quadratic power dependence on $\underline{v}_{ij} \cdot \underline{r}_{ij}/|\underline{r}_{ij}|$ is analogue to both the linear and the quadratic components of the artificial viscosity terms (15). The linear term $\propto c_{s} v_{shock}$ is based on the viscosity of a gas. The quadratic term $\propto (c_{s} v_{shock})^{2}$ (Von Neumann - Richtmyer - like) is necessary to handle strong shocks. These contributions involve a dissipative power, whose effect corresponds to an increase of the gas pressure. Therefore, we adopt the formulation (eq. 21) as $p_{i}^{\star}$ and $p_{j}^{\star}$ in the SPH formulation of the momentum (eq. 9) and energy equations (eqs. 10 or 11):

\begin{equation}
\frac{d \underline{v}_{i}}{dt} = - \sum_{j=1}^{N} m_{j} 
\left( \frac{p_{i}^{\star}}{\rho_{i}^{2}} + \frac{p_{j}^{\star}}{\rho_{j}^{2}} \right) \nabla_{i} W_{ij}
\end{equation}

\begin{equation}
\frac{d \epsilon_{i}}{dt} = \frac{1}{2} \sum_{j=1}^{N} m_{j} \left( \frac{p_{i}^{\star} }{\rho_{i}^{2}} + \frac{p_{j}^{\star}}{\rho_{j}^{2}}\right) \underline{v}_{ij} \cdot \nabla_{i} W_{ij},
\end{equation}

\begin{equation}
\frac{d}{dt} E_{i} = - \sum_{j=1}^{N} m_{j} \left( \frac{p_{i}^{\star} \underline{v}_{i}}{\rho_{i}^{2}} + \frac{p_{j}^{\star} \underline{v}_{j}}{\rho_{j}^{2}} \right) \cdot \nabla_{i} W_{ij}.
\end{equation}

  This simple reformulation, allows us to to keep the same Courant - Friedrichs - Lewy condition as for the timestep computation, substituting only the sound velocity $c_{s}$ with $c_{s} - v_{shock}$.

%-------------------------------------------------- FIG 1 START
\begin{figure}
\resizebox{\hsize}{!}{\includegraphics[clip=true]{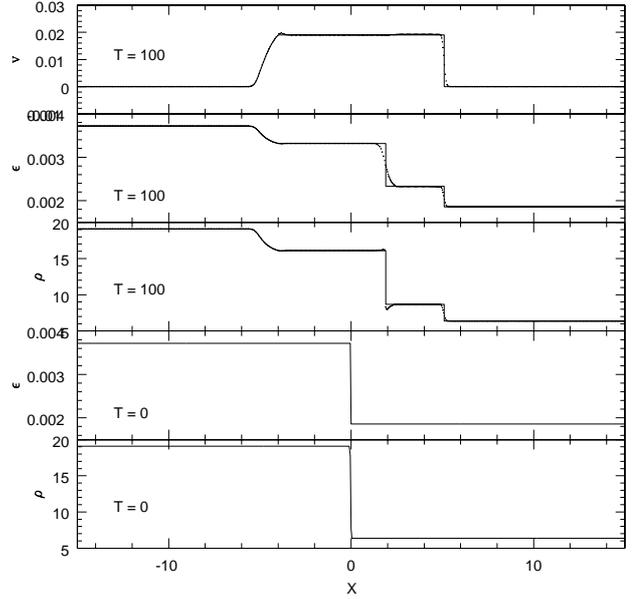}}
\caption{1D shock tube tests as far as both analytical (solid line) and our SPH-Riemann (dots) results are concerned. Density $\rho$, thermal energy $\epsilon$ and velocity $v$ are plotted at time $T = 100$. Density and thermal energy of particles initially at rest at time $T = 0$ are also reported. The initial velocity is zero throughout.}
\end{figure}
%-------------------------------------------------- FIG 1 END

\section{1D Sod shock tube tests}

  In this section a comparison of analytical and our SPH - Riemann 1D shock tube test results (Sod 1978), also with the initial particle configuration (time $T = 0$), is made. Notice that the so called analytical solution of the Riemann problem is obtained through iterative procedures left-right, applying to the discontinuity the Rankine - Hugoniot "jump" solution. Figg. 1 and 2 shows results concerning the particle density, thermal energy per unit mass and velocity at the same final computational time ($T = 100$). Throughout our SPH-Riemann simulations, the initial particle resolution length is $h = 5 \cdot 10^{-2}$. The whole computational domain is built up with $2001$ particles from $X = 0$ to $X = 100$, whose mass is different, according to the shock initial position. At time $T = 0$ all particles are motionless. $\gamma = 5/3$, while the ratios $\rho_{1}/\rho_{2} = 3$ and $\epsilon_{1}/\epsilon_{2} = 2$ in Fig. 1, and  $\rho_{1}/\rho_{2} = 3$ and $\epsilon_{1}/\epsilon_{2} = 1$ in Fig. 2, between the two sides left-right. The first $5$ and the last $5$ particles of the 1D computational domain, keep fixed positions and do not move. The choice of the final computational time is totally arbitrary, since the shock progresses in time.

%-------------------------------------------------- FIG 2 START
\begin{figure}
\resizebox{\hsize}{!}{\includegraphics[clip=true]{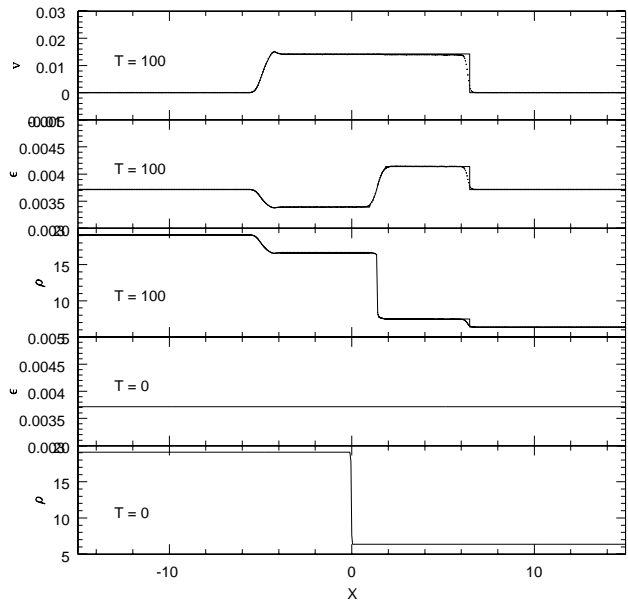}}
\caption{The same  as in Fig. 1. In this example, the initial discontinuity does not affect the thermal energy per unit mass, being the gas initially isothermal, while the initial velocity is zero throughout.}
\end{figure}
%-------------------------------------------------- FIG 2 END

  Results, where our SPH-Riemann solution is applied to SPH, are in a good comparison with the analytical solution. Discrepancies involve only 4 particles at most. This means that, according to the cautionary remarks on \S 2.3, the physical dissipation introduced in the EoS (eqs. 37, 38) is effective. Any artificial viscosity contribution is totally absent, as well as no thermodynamic properties of neighbour particle are involved, as the application of Godunov - type solver (Yukawa 1997; Inutsuka 2002; Parshikov \& Medin 2002; Molteni \& Bilello 2003) does.

\section{Discussion and conclusions}

  The comparison of 1D Sod shock tube analytical results with SPH - Riemann ones, where a revision of the EoS within the Riemann problem is made, are quite successful. In particular, in our modelling, neither a parametrized artificial viscosity, nor any dependence on spatial resolution length $h$, nor a Godunov solver, nor additional computational time are involved. However, to give a generalization, we need only one general EoS and not a separation of the EoS according to the kinematic of the flow. To this purpose, we can generalize the EoS: $p^{\star} = \rho c^{2}_{s} (1 - v_{shock}/c_{s})^{2}/\gamma$ as:

\begin{equation}
p^{\star} = \frac{\rho}{\gamma} c_{s}^{2} \left(1 - C \frac{v_{R}}{c_{s}} \right)^{2},
\end{equation}

where $C \rightarrow 1$ for $v_{R} = \underline{v}_{ij} \cdot \underline{r}_{ij}/|\underline{r}_{ij}| < 0$, whilst $C \rightarrow 0$ otherwise. A simple empirical formulation can be,

\begin{equation}
C = \frac{1}{\pi}  \textrm{arccot} \left( R \frac{v_{R}}{c_{s}} \right),
\end{equation}

where $R \gg 1$. $R$ is a large number describing how much the flow description corresponds to that of an ideal gas. To this purpose, $R \approx \lambda/d$, being $\lambda \propto \rho^{-1/3}$ the molecular mean free path, and $d$ the mean linear dimension of gas molecules. Further, improvements can be made next future. The comparison with analytical results of 1D Sod shock tubes in the same test in \S 4 gives us the same results showed in Figg. 1 and 2. The simple EoS for inviscid ideal gases: $p = (\gamma - 1) \rho \epsilon$ cannot be strictly applied in shock problems by the fact that, solving the Euler equations, not only instabilities and spurious heating come out, but also that this empirical EoS derived by Charles, Volta Gay - Lussac and Boyle - Mariotte laws, should be strictly applied only either in equilibrium configurations or to "quasi-static" evolutions of thermodynamic systems without any dissipation. This is a restriction that does not match with shock flow dynamics, when dissipation on the shock front cannot be neglected. Those techniques involving Godunov - type schemes also introduce some dissipation mechanisms (Park \& Kwon 2003). Therefore, according to these schemes, the solution of Euler equations is made possible because of dissipative mechanisms introduced in the numerical methods.

  To conclude, although built up on empirical basis, the general EoS here proposed, shows the correct behaviour, even in the presence of dissipative shocks.

\clearpage

\end{document}